\documentclass[showpacs,twocolumn,floatfix,superscriptaddress]{revtex4}
\usepackage{graphicx}

\newcommand{\bq}{\begin{equation}}
\newcommand{\ee}{\end{equation}}
\newcommand{\fr}[2]{\frac{#1}{#2}}

\begin{document}

\title{ 
Hypersensitivity to perturbations of quantum-chaotic wave-packet dynamics}
\author{P. G. Silvestrov}
\affiliation{Instituut-Lorentz, Universiteit Leiden, P.O. Box 9506, 2300 RA
Leiden, The Netherlands}
\affiliation{Budker Institute of Nuclear Physics, 630090 Novosibirsk, Russia}
\author{J. Tworzyd\l o}
\affiliation{Instituut-Lorentz, Universiteit Leiden, P.O. Box 9506, 2300 RA
Leiden, The Netherlands}
\affiliation{Institute of Theoretical Physics, Warsaw University,
Hoza 69, 00-681 Warszawa, Poland}
\author{C. W. J. Beenakker}
\affiliation{Instituut-Lorentz, Universiteit Leiden, P.O. Box 9506, 2300 RA
Leiden, The Netherlands}

\date{30 June 2002}
\begin{abstract}

We re-examine the problem of the ``Loschmidt echo'', which measures
the sensitivity to perturbation of quantum chaotic dynamics. The overlap 
squared $M(t)$ of two wave packets evolving under slightly different 
Hamiltonians is shown to have the double-exponential initial decay
$\propto \exp(-{\rm constant}\times e^{2\lambda_0 t})$ in the main part of 
phase space. The coefficient $\lambda_0$ is the self-averaging 
Lyapunov exponent. The average decay $\overline{M}\propto e^{-\lambda_1 t}$ is
single exponential with a different coefficient $\lambda_1$. The
volume of phase space that contributes to $\overline{M}$ vanishes
in the classical limit $\hbar\rightarrow 0$ for times less than the 
Ehrenfest time $\tau_E=\fr{1}{2}\lambda_0^{-1}|\ln \hbar|$.
It is only after the Ehrenfest time that the average decay is 
representative for a typical initial condition.

\end{abstract}

\maketitle

Chaos in classical mechanics is characterised by an exponential
sensitivity to initial conditions: The separation of two trajectories that
are initially close together increases in time $\propto e^{\lambda t}$ with a
rate given by the Lyapunov exponent $\lambda$. There is no 
such sensitivity in quantum mechanics, because the overlap of two 
wave functions is time independent. This
elementary observation is at the origin of a large literature (reviewed in a
textbook~\cite{Haake}) on quantum characterisations of chaotic dynamics.

One particularly fruitful line of research goes back to the proposal of
Schack and Caves~\cite{SchackCaves}, motivated by earlier work of
Peres~\cite{Peres}, to characterise chaos by the sensitivity to
perturbations. Indeed, if one and the same state $\psi_0$ evolves under
the action of two different Hamiltonians $H$ and $H+\delta H$, then
the overlap
\bq\label{loschm}
M(t)=|\langle\psi_0|e^{i(H+\delta H)t/\hbar}e^{-iHt/\hbar}
|\psi_0\rangle|^2
\ee
is not constrained by unitarity. Jalabert and Pastawski\cite{Jalabert}
discovered that $M(t)$ (which they referred to as the ``Loschmidt echo'') 
decays $\propto e^{-\lambda t}$ if $\psi_0$ is a
narrow wave packet in a chaotic region of phase space, providing an
appealing connection between classical and quantum chaos.

The discovery of Jalabert and Pastawski gave a new 
impetus~\cite{Gm} to
what Schack and Caves called ``hypersensitivity to perturbations'' of quantum
chaotic dynamics. The present paper differs from this body of literature
in that we consider the {\em statistics} of $M(t)$ as $\psi_0$ varies over
the chaotic phase space. We find that the average decay ${\overline M}
(t)\propto e^{-\lambda t}$ is due to regions of phase space that become 
vanishingly small in the classical limit $\hbar_{\rm eff}\rightarrow 0$. 
(The effective Planck constant $\hbar_{\rm eff}=\hbar/S_0$ is set by the 
inverse of a typical action $S_0$.) 
%We will discuss in details the case of 
%chaotic quantum map, but our results are easily generalised for any 
%classically chaotic system provided that $\hbar_{\rm eff}\ll 1$.) 
The dominant decay is a {\em double}
exponential $\propto \exp(-{\rm constant}\times e^{2\lambda t})$, so it is 
truly ``hypersensitive''. The slower single exponential decay is recovered
at the Ehrenfest time $\tau_E=\fr{1}{2}\lambda^{-1}|\ln \hbar_{\rm eff}|$.

Before presenting our analytical theory we show in Fig.~1 the data 
from a numerical 
simulation that illustrates the hypersensitivity mentioned above.
The Hamiltonian is the quantum kicked rotator~\cite{Haake}
\bq\label{rotat}
H=\fr{\hat{p}^2}{2} +K\cos x\sum_n\delta(t-n) \ ,
\ \hat{p}=\fr{\hbar}{i} \fr{d}{dx} .
\ee
The perturbed Hamiltonian $H'=H+\delta H$ is obtained by the replacement
$K\rightarrow K+\delta K$. The coordinate $x$ is periodic,
$x\equiv x+2\pi$. To work with a finite dimensional Hilbert space
we discretize $x_k=2\pi k/N$, $k=1,2,...N$.
The momentum $p_m=m\hbar$ is a multiple of $\hbar$, to ensure
single-valued wavefunctions. For  $\hbar\equiv\hbar_{\rm eff}=2\pi/N$
the restriction to the first Brillouin zone results in a single band
$p_m=2\pi m/N$, $m=1,2,...N$.
The time evolution $e^{-iHn/\hbar}\equiv U^n$ 
after $n$ periods, 
of the initial Gaussian wave packet $\psi_k=N^{-1/2} \exp\big(
\pi N^{-1}[2im_0k-(k-k_0)^2]
\big)$,
is given by the Floquet operator in the
$x$-representation:
\bq\label{Floquet}
U_{k'k}=
\fr{1}{\sqrt{N}}
\exp \left(
\fr{i\pi(k'-k)^2}{N}
-i\fr{NK}{2\pi}\cos 
\fr{2\pi k}{N}
\right)
.
\ee
We use the fast Fourier transform algorithm to compute $U^n$
for $N$ up to $10^6$~\cite{Ketzmerick}. 

We study the statistics of
$M(t)$ by comparing in Fig.~1 three different ways of
averaging over initial positions $(m_0,k_0)$ of the Gaussian wave packet. 
We used $K=10$, $\delta K= 1.6 \cdot 10^{-3}$ and $N=10^6$ 
($\hbar_{\rm eff}=6.28 \cdot 10^{-6}$).
While the average $\overline{M}$ decays exponentially, the two
logarithmic averages have a much more rapid initial decay.
We estimate that $M<10^{-23}$ at $n=3$
for about $30\%$ of randomly chosen initial conditions.
For the same point $n=3$ only $9\%$ of initial conditions
(corresponding to $M>0.2$) account for $80\%$ of the total value of 
$\overline{M}$. The typical decay of $M(t)$ is therefore much
more rapid than the exponential decay of the average $\overline{M}$.

Statistical fluctuations also affect the decay rate of $\overline{M}$,
set by the Lyapunov exponent according to Ref.~\cite{Jalabert}.
The definition of the Lyapunov exponent $\lambda_0= 
\lim_{t\rightarrow\infty} t^{-1}\ln|\delta x(t)/\delta x(0)|$ 
gives $\lambda_0=1.65$
for the classical kicked rotator with $K=10$.
However, $\overline{M(t)}$ in Fig.~1 has
exponent $\lambda_1=1.1$, defined by
\bq\label{lambdan}
\lambda_j =-\lim_{t\rightarrow\infty} 
(jt)^{-1}\ln \overline{|\delta x(t)/\delta x(0)|^{-j}}.
\ee
Since fluctuations of $t^{-1}\ln|\delta x(t)/\delta x(0)|$
decrease like $t^{-1/2}$ the Lyapunov exponent $\lambda_0$
is self-averaging~\cite{Henning}, while the $\lambda_j$'s are not.

\begin{figure}[t]
\includegraphics[width=8.5cm]{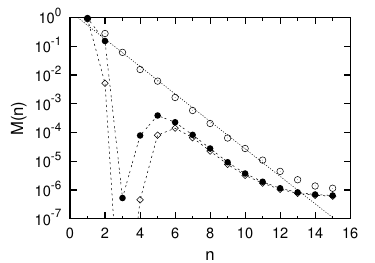}
\caption{
The overlap $M$ at $t=n$ for the quantum kicked rotator for three
different ways of averaging: 
$\circ$ $\overline{M}$,
$\diamond$ $\exp(\overline{\ln M})$,
$\bullet$ $\exp[-\exp(\overline{\ln(-\ln M)})]$. 
We took $K=10$, $\delta K=1.6 \cdot 10^{-3}$, $N= 10^6$. 
Averages are taken over
$2000$ random initial conditions of a Gaussian wave packet.
The dotted line shows the Lyapunov decay $\propto e^{-n\lambda_1}$ with 
$\lambda_1=1.1$. At $n=3$ we have only an upper bound for the 
logarithmic averages because cancelations in the
calculation limit the accuracy.
}
\end{figure}

For an analytical description we start from the Gaussian 
one-dimensional wave packet 
\bq\label{pack} 
\psi(x)=\left( 
\fr{\alpha}{\pi\hbar} 
\right)^{1/4} 
\exp \left( 
i\fr{p_0 x}{\hbar}+(i\beta-\alpha)\fr{(x-x_0)^2}{2\hbar} 
\right) .  
\ee 
The wave packet is centered at the point $x_0(t), p_0(t)$ which moves 
along a classical trajectory.
Initially $\beta(t=0)=0$ and $\alpha(t=0)=1$.  
Divergence of trajectories leads to the exponential 
broadening of the packet, thus 
$\alpha(t)\propto\exp(-2\lambda t)$. Since
$\alpha\ll 1$ for $t\gg 1/\lambda$, the wave packet in phase space 
becomes highly elongated with length 
$l_\parallel=\sqrt{\hbar (1+\beta^2)/\alpha}$ 
and width $l_\perp =\hbar/l_\parallel$. The parameter $\beta=
\Delta p/\Delta x$ represents the tilt angle of the elongated wave
packet~\cite{endnote}. The Gaussian  approximation~(\ref{pack})
breaks down
at the Ehrenfest time $\tau_E=\frac{1}{2}\lambda^{-1}|\ln\hbar|$,
when $l_\parallel$ becomes of the order of the size of 
the system.

We assume that $\psi$ evolves according to Hamiltonian $H(K)$ and
$\psi'$ according to $H'=H(K+\delta K)$. 
The overlap 
$M=|\langle\psi'|\psi\rangle|^2$ 
of the two Gaussian wave packets 
is 
\bq\label{res}
M=
\sqrt{
\fr{\alpha\alpha'}
{\overline{\alpha}^2+4\delta\beta^2}
}
\exp\left(
-\fr{\overline{\alpha}(\delta p-{\overline \beta}\delta x)^2}
{2(\overline{\alpha}^2+4\delta\beta^2)\hbar}
-\fr{\alpha\alpha'
\delta x^2}{2\overline{\alpha}\hbar}
\right),
\ee
in terms of the (weighted) mean $\overline{\alpha}=(\alpha+\alpha')/2$, 
${\overline \beta}=(\beta'\alpha +\beta\alpha')/(\alpha+\alpha')$ 
and difference $\delta p=p_0'-p_0$, $\delta x=x_0'-x_0$,  
$\delta \alpha=\alpha'-\alpha$, $\delta \beta=\beta'-\beta$.
In order of magnitude,
$\delta\beta/\overline{\beta}\simeq\delta\alpha/\overline{\alpha}\simeq\delta
K\ll 1$. The displacement vector $(\delta x,\delta p)$ has component
$\Delta_{\parallel}\simeq\delta K e^{\lambda t}$ parallel to the elongated 
wave packets and component $\Delta_{\perp}\simeq\delta K$ perpendicular 
to them (see Fig.~2). 

Depending on the strength of perturbation 
one may distinguish three main regimes:
$\delta K<{\hbar}$, $\hbar <\delta K< \sqrt{\hbar}$, and
$\delta K > \sqrt{\hbar}$. We will consider in detail the 
intermediate regime $\hbar <\delta K< \sqrt{\hbar}$ and discuss the
two other regimes more briefly at the end of the paper. 
(The simulations of Fig.~1 are at the upper end of the intermediate regime,
since $\delta K=1.6 \cdot 10^{-3}$ and $\sqrt{\hbar}=2.5 \cdot 10^{-3}$.)
The three regimes may be characterised by the relative magnitude
of the Ehrenfest time $\tau_E$ and
the perturbation dependent time scale $\tau_0=
\fr{1}{2}\lambda^{-1}|\ln\delta K|$. 
In the intermediate regime one has $\fr{1}{2}\tau_E<\tau_0<\tau_E$.

To estimate the relative magnitude of the two terms in the exponent of 
Eq.~(\ref{res}) we write
\begin{eqnarray}
&&\delta p-\overline{\beta}\delta
x =(1+\overline{\beta}^{2})^{1/2}\Delta_{\perp}=f\delta
K,\label{deltapdeltax}\\
&&\frac{\overline{\alpha}}{\overline{\alpha}^{2}+4\delta\beta^{2}}\equiv
Q=\frac{e^{2\lambda t}}{1+(ge^{2\lambda t}\delta K)^{2}}.\label{alphabeta}
\end{eqnarray}
Here $f$ and $g$ are functions of order unity of time $t$ and the initial
location $x_{i},p_{i}$ of the wave packet. The second term in the 
exponent~(\ref{res}) is
of order $\overline{\alpha}\delta x^{2}/\hbar\simeq\delta K^{2}/\hbar$, while
the first term is of order $Q\delta K^{2}/\hbar$. Since $Q\gg 1$ for
$t<2\tau_0$, and since $2\tau_0>\tau_E$ in the intermediate regime, 
we may neglect the second term relative to the
first term within the entire range $t<\tau_{E}$ of validity of the Gaussian
approximation. Eq.\ (\ref{res}) thus simplifies to
\begin{equation}\label{MQ}
M=(\overline{\alpha}Q)^{1/2}\exp[-\case{1}{2}Q(f\delta
K)^{2}/\hbar].\label{Msimple}
\end{equation}

We seek the statistics of $M(t)$ generated by varying $x_{i},p_{i}$. 
The statistics is non-trivial because
fluctuations in $f$ of order unity cause exponentially large fluctuations in
$M$ if $Q\delta K^{2}/\hbar\gg 1$, which is the case for $2\tau_0-\tau_E
<t<\tau_E$. The average of $M$ is then
dominated by the nodal lines $x_{n}(p)$ in phase space at which $f$ vanishes
(at a particular time $t$). If $\Delta x_{n}$ is the typical spacing of these
lines at constant $p$, then the derivative $\partial f/\partial x_{i}$ at
$x_{n}$ is of order $1/\Delta x_{n}$. 
This yields 
\begin{eqnarray}\label{Maverage}
\overline{M}&=&(\overline{\alpha}Q)^{1/2}\int \frac{dx}{\Delta
x_{n}}\,\exp\left[-\left(\frac{x-x_{n}}{\Delta x_{n}}\right)^{2}\frac{Q\delta
K^{2}}{2\hbar}\right]\nonumber\\
&\simeq&{(\overline{\alpha}\hbar/\delta K^{2})^{1/2}}
=
({\sqrt{\hbar}}/{\delta K})
e^{-\lambda t}.
\end{eqnarray}
Assuming independent fluctuations in the (perturbation dependent) 
distribution of nodal lines and 
in the rate of divergence
of trajectories for the individual Hamiltonian, we
incorporate fluctuations in $\lambda$ in Eq.~(\ref{Maverage})
via
$\overline{\exp(-\lambda t)}\rightarrow 
\exp(-\lambda_1 t)$, in accordance with~(\ref{lambdan}). 
Hence we recover the exponential decay of the Loschmidt echo \cite{Jalabert},
although with the exponent $\lambda_1$ instead of $\lambda_0$
(in agreement with the numerics of Fig.~1).
The exponential decay sets in for $t>2\tau_0-\tau_E$, while for shorter times
$\overline{M}$ remains close to unity~\cite{Casati}.

The volume ${\cal V}$ of phase space near the nodal lines contributing to
$\overline{M}$ is of order ${\cal V}=(\hbar/Q\delta K^{2})^{1/2}$. This volume
decreases exponentially in time for
$t<\tau_{0}$, reaching the minimal
value ${\cal V}_{0}=\sqrt{\hbar/\delta K}\ll 1$ at $\tau_{0}$. 
For larger times
${\cal V}$ increases, saturating at a value of order unity at $\tau_{E}$. We
therefore conclude that the average $\overline{M}$ is only representative for
the typical decay if $t>\tau_{E}$. For smaller times the average is dominated
by rare fluctuations that represent only a small fraction of the chaotic phase
space.

To obtain an average quantity that is representative for a typical point in
phase space we take logarithmic averages of Eq.~(\ref{MQ}).
For $t<\tau_0$ one has 
\bq
\overline{\ln M}\simeq -
(\delta K^{2}/\hbar)\exp(2\lambda_{-2}
t),
\ee
\begin{equation}\label{lnlnMshorttime}
\overline{\ln\ln(1/M)}=2\lambda_{0}t-\ln(\hbar/\delta K^{2})+{\cal
O}(1).
\end{equation}
(The coefficient $\lambda_{-2}$ in $\overline{\ln M}$ appears because 
we average the square of displacement.)
The double logarithmic average (\ref{lnlnMshorttime}),
given by the self-averaging Lyapunov exponent $\lambda_{0}$, is least 
sensitive 
to fluctuations and is representative for the main part of phase space.
The typical overlap thus has the double-exponential decay
\begin{equation}\label{Mdecayshorttime}
M\simeq\exp\big(-{\rm constant}\times(\delta K^{2}/\hbar)e^{2\lambda_{0}
t}\big),
\end{equation}
down to a minimal value $M_{0}\simeq\exp(-\delta K/\hbar)$ at $t=\tau_{0}$.

The initial decay (\ref{Mdecayshorttime}) for $t\ll\tau_0$ is the same as 
obtained in Ref.~\cite{Prosen} for the classical fidelity
(defined as the overlap of two classical phase space densities). In that
problem the role of $\hbar$ is played by the initially occupied volume
of phase space. A superexponential decay of the classical fidelity has 
also been obtained by Eckhardt~\cite{Eckhardt}.

\begin{figure}[t]
\includegraphics[width=8cm]{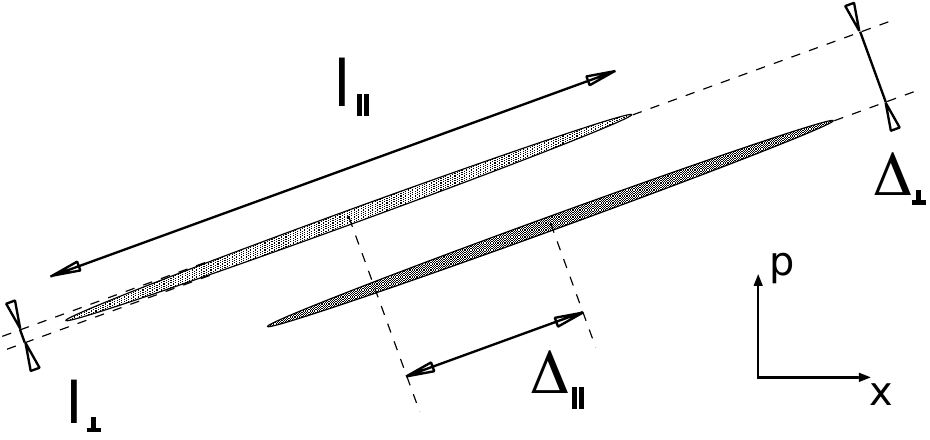}
\caption{
Schematic illustration of two perturbed wave packets in phase space
for $t<\tau_0$.
}
\end{figure}

The origin of the decay~(\ref{Mdecayshorttime}) is illustrated in Fig.~2. 
For $t<\tau_{0}$
the wave packets are nearly parallel ($\delta\beta\ll\overline{\alpha}$),
displaced laterally by an amount $\Delta_{\perp}\propto\delta K$. 
Their overlap
is an exponential function $\propto\exp(-\Delta_{\perp}^{2}/l_{\perp}^{2})$,
where the width $l_{\perp}$ of each wave packet decreases exponentially 
in time
$\propto e^{-\lambda t}$. Hence the double-exponential decay.

For $t>\tau_{0}$ (when $\delta\beta\gg\overline{\alpha}$) the overlap of the
two wave packets is dominated by their crossing point $x_{c},p_{c}$. The
overlap $M\simeq\exp(-{\rm constant} 
\times|x_{c}-x_{0}|^{2}/l_{\parallel}^{2})$
now {\em increases\/} with time, because $l_{\parallel}\propto e^{\lambda t}$.
Since $|x_{c}-x_{0}|\simeq\Delta_{\perp}/\delta\beta\simeq f$, the crossing
point falls outside of the range of validity of the Gaussian approximation
unless $|f|\ll 1$. The result (\ref{Maverage}) is justified (because it is
dominated by nodes of $f$), but we can not use the Gaussian approximation to
extend the formula (\ref{Mdecayshorttime}) for the typical decay to
$t>\tau_{0}$. The typical decay
and the average decay become the same at $\tau_{E}$, so the typical $M$ should
increase from its minimal value $M_{0}$ at $\tau_{0}$ to the value
$M_{E}=(\sqrt{\hbar}/\delta K)e^{-\lambda\tau_{E}}=\hbar/\delta K$ at
$\tau_{E}$. Both $M_{0}$ and $M_{E}$ are $\ll 1$, but $M_{0}$ is exponentially
small in $\delta K/\hbar$ while $M_{E}$ is only algebraically small.

\begin{figure}[t]
\includegraphics[width=6.3cm]{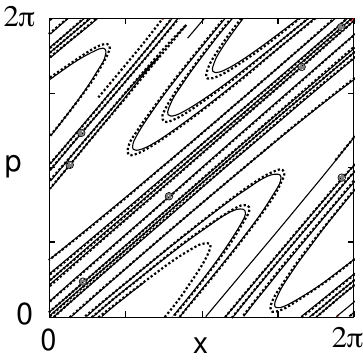}
\caption{
Two perturbed wave packets in phase space for $\tau_E<t<\tau_E+2\tau_0$.
The lines show $p(x)$ (solid) and $p'(x)$ (dashed) extracted from the 
Husimi function 
evolved with the quantum kicked rotator,
for $N=10^6$, $K=7$, $\delta K=0.1$, $n=5$. Dots show the crossing points
$x_j$ that contribute to the overlap in stationary phase approximation.
}
\end{figure}

For $t>\tau_{E}$ one
can use the semiclassical 
WKB description of elongated wave packets, along the lines of
Refs.~\cite{BerryB}.
The phase space representation of the wave function $\psi$ is concentrated
along the line on the torus $p(x)$ of length 
$l_\parallel\simeq\sqrt{\hbar}e^{\lambda t}\gg 1$, see Fig.~3.
The function $p(x)$ is multivalued and each branch $k$ has a WKB wave function
with amplitude $\rho_k\approx 1/l_\parallel$ and phase $\sigma_k$:
\bq\label{semiGM}
\psi=\sum_k \sqrt{\rho_k} e^{i\sigma_k/\hbar} \ , \ p_k=d\sigma_k/dx .
\ee
For $\delta K>\hbar$ the overlap of two oscillating wave 
functions $\psi$,$\psi'$ of the form (\ref{semiGM})
may be found in stationary phase approximation.
The stationary points are given by the crossings $p(x_j)=p'(x_j)$ of
the two lines $p(x)$, $p'(x)$ given by the evolution with Hamiltonians
$H,H'$. 
For  $\tau_E<t<\tau_E+2\tau_0$ the number of crossings $N_c$ 
is proportional to $l_\parallel$ and independent of $\delta K$.
This is because both the lateral displacement of $p$ and $p'$ and
their relative angle are of the same order $\delta K$.
(In Fig.~3 we have ${l_\parallel}\simeq 20N_c$.)
Each crossing contributes to $\langle \psi|\psi'\rangle$ an amount
\begin{eqnarray}\label{Pc}
P_j&=&\sqrt{\rho(x_j)\rho'(x_j)}\int dx 
\exp\left[
i\fr{\kappa(x-x_j)^2}{2\hbar}+i\phi_j
\right]
\nonumber\\
&\simeq&
({e^{i\phi_j}}/{l_\parallel})\sqrt{{\hbar}/{\delta K}},
\end{eqnarray}
where $\kappa={d^2(\sigma-\sigma')}/{dx^2}|_{x_j}
\simeq\delta K$ and $\hbar\phi_j=\sigma(x_j)-\sigma'(x_j)$.
The phase $\phi_j$ varies randomly from one crossing to the other, leading to
\bq\label{sc}
\overline{M}=\fr{\hbar}{l_\parallel^2\delta K}
\sum_{j,j'=1}^{N_c}{\overline{
e^{i(\phi_j-\phi_{j'})}
}} 
\simeq \fr{\hbar}{l_\parallel\delta K}
\simeq\fr{\sqrt{\hbar}}{\delta K}e^{-\lambda_1 t}.
\ee
Because of the large number 
of crossings
there is now little difference between $\overline{M}$ 
and logarithmic averages.
For $t>\tau_E+2\tau_0$
the number of crossings 
becomes 
$N_c\simeq \delta K l_\parallel^2$. (The distance between almost parallel
segments of $p'(x)$ is of order $1/l_\parallel$, and the line $p(x)$
crosses at the angle $ \delta K$ about $\delta K l_\parallel\gg 1$ 
segments per unit length.)
This leads to saturation of the overlap at
$\overline{M} \simeq\hbar$.

\begin{figure}[t]
\includegraphics[width=8.5cm]{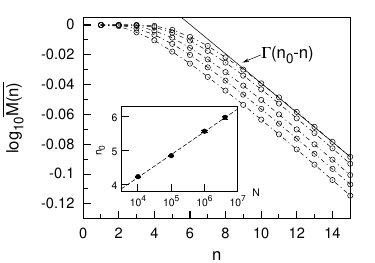}
\caption{ Decay of the average overlap for the quantum kicked rotator 
($K=10$) in the golden-rule regime.
We keep $\Gamma = 0.023(N\delta K)^2$ fixed 
by taking $\delta K\equiv 1/N$.
Circles from bottom to top give $\overline{M}$ for
$N=10^3,10^4,10^5,10^6,4\cdot 10^6$.
The inset demonstrates that
$n_0$ scales with $\log N$,
as expected for the Ehrenfest time.}
\end{figure}

This completes our discussion of the intermediate regime
$\hbar < \delta K < \sqrt{\hbar}$. 
We conclude with a brief discussion of the two other regimes. 
For $\delta K > \sqrt{\hbar}$ the longitudinal displacement of the
packets exceeds their lengths $\Delta_\parallel> l_\parallel$.
The logarithmic averages now remain the same,
but  $\overline{M}$ is changed. The dominant contributions
to $\overline{M}$ are now given by the rare events for which both
$\Delta_\perp$ and $\Delta_\parallel$ vanish.
This leads to $\overline{M}\simeq
(\hbar/\delta K^2)e^{-\lambda_1 t}$ for $t<2\tau_0$. (The same Lyapunov 
decay as in the intermediate regime, but with a much smaller 
prefactor.) For $2\tau_0<t<\tau_E$
the length of each wave packet remains small,
$l_\parallel\simeq \sqrt{\hbar} e^{\lambda t}\ll 1$, but the displacement
saturates at the maximal value $\Delta_\parallel \simeq 1$.
In this time range the average overlap has a plateau at $\overline{M}\simeq
{\hbar/\delta K}$. Finally, for $t>\tau_E$ the decay~(\ref{sc}) 
$\overline{M}\simeq
(\sqrt{\hbar/}\delta K)e^{-\lambda_1 t}$ is recovered.

In the remaining regime $\delta K\ll \hbar$ we find from Eq.~(\ref{res})
that $M(t)$ remains close to unity for $t<\tau_E$, regardless of the 
initial location of the wave packet. (This also results in insensitivity
to the way of averaging.) The golden-rule decay~\cite{Gm},
with rate $\Gamma\simeq (\delta K/\hbar)^2$, sets in only after the
Ehrenfest time: $\overline{M}\simeq \exp[-\Gamma(t-\tau_E)]$ for $t>\tau_E$.
These results are depicted in Fig.~4. 
Golden-rule decay persists until the Heisenberg time $t_H\simeq 
1/\hbar$ or the saturation time $\Gamma^{-1}|\ln\hbar|$,
whichever is smaller. (Only the initial decay is shown in Fig.~4.)
The Gaussian decay~\cite{Gm} sets in for $t>t_H$ provided that 
$\delta K<\hbar^{3/2}$.

In summary, we have shown that statistical fluctuations play a dominant role 
in the problem of the Loschmidt echo on time scales below the Ehrenfest 
time. While the decay of the squared overlap $M(t)$ of two perturbed wave
packets is exponential on average, as obtained previously~\cite{Jalabert},
the typical decay is double exponential. It is only after the Ehrenfest
time that the main part of phase space follows the single-exponential decay 
of $\overline{M}$. The Ehrenfest time has been heavily studied in
connection with the quantum-to-classical correspondence~\cite{Gm}.
The role which this time scale plays in suppressing statistical
fluctuations 
has not been anticipated in this large body of literature.

We acknowledge  discussions with E.~Bogomolny.
This work 
was supported by the Dutch Science Foundation NWO/FOM.
J.T. acknowledges support of the European 
Community's Human Potential Program under contract 
HPRN--CT--2000-00144, Nanoscale Dynamics.

\end{document}